# Perfect Spatiotemporal Optical Vortices

**H**AIHAO **F**AN[1,†], **Q**IAN **C**AO[1,2,†], **X**IN **L**IU[3,4,†], **A**NDY **C**HONG[5,6], **Q**IWEN **Z**HAN[1,2,7,*]

[1] *School of Optical-Electrical and Computer Engineering, University of Shanghai for Science and Technology, Shanghai 200093, China.*

[2] *Zhangjiang Laboratory, Shanghai 201210, China*

[3] *Shandong Provincial Engineering and Technical Center of Light Manipulations and Shandong Provincial Key Laboratory of Optics and Photonic Device, School of Physics and Electronics, Shandong Normal University, Jinan 250358, China.*

[4] *Collaborative Innovation Center of Light Manipulations and Applications, Shandong Normal University, Jinan 250358, China*

[5] *Department of Physics, Pusan National University, Busan 46241, Republic of Korea*

[6] *Institute for Future Earth, Pusan National University, Busan 46241, Republic of Korea*

[7] *Westlake Institute for Optoelectronics, Fuyang, Hangzhou, China*

[†]These authors contributed equally to this work.

*qwzhan@usst.edu.cn

**Abstract:** Recently, spatiotemporal optical vortices (STOVs) with transverse orbital angular momentum have emerged as a significant research topic. While various STOV fields have been explored, they often suffer from a critical limitation: the spatial and temporal dimentions of the STOV wavepacket are strongly correlated with the topological charge. This dependence hinders the simultaneous achievement of high spatial accuracy and high topological charge. To address this limitation, we theoretically and experimentally investigate a new class of STOV wavepackets generated through the spatiotemporal Fourier transform of polychromatic Bessel-Gaussian beams, which we term as perfect spatiotemporal optical vortices. Unlike conventional STOVs, perfect STOVs exhibit spatial and temporal diameters that are independent of the topological charge. Furthermore, we demonstrate the generation of spatiotemporal optical vortex lattices by colliding perfect STOV wavepackets, enabling flexible manipulation of the number and sign of sub-vortices.

## 1. Introduction

An optical vortex, characterized by a helical phase structure of the form $\exp(il\theta)$, features a central phase singularity and a hollow intensity profile. The topological charge, denoted by $l$, quantifies the number of phase twists accumulated during one wavelength of propagation [1-3]. In the early1990s, Allen and colleagues demonstrated that the longitudinal orbital angular momentum (OAM) of a photon is inherently carried by a spatial vortex beam, with the magnitude of the OAM being proportional to its topological charge [4,5]. Since then, optical vortex beams carrying longitudinal OAM, including Laguerre-Gaussian beams, Bessel-Gaussian beams, and others, have been extensively demonstrated and applied in a wide range of applications, such as light-matter interactions [6,7], optical imaging [8,9], quantum information[10,11], and optical communications[12,13], as well as nonlinear optics[14,15].

Recently, spatiotemporal optical vortices (STOVs) defined in the space-time domain have garnered increasing attention due to their ability to carry unique transverse OAM of light [16,17]. The development of STOVs can be traced back to early predictions [18,19] and has recently been experimentally demonstrated through both nonlinear [20] and linear methods [21,22]. The successful generation of STOV wavepackets has catalyzed further exploration into spatiotemporal topology[23,24], harmonic generations [25,26,27], and other forms of matter waves [28]. This progress has led to the emergence of novel spatiotemporal optical wavepackets [16,17], including spatiotemporal Bessel [29,30], crystal [31], Laguerre and Hermite Gaussian wavepackets[32]. However, the beam radii of these conventional optical vortices—whether spatial vortex beams or STOV wavepackets — strongly depends on their topological charge number, which can hinder their applications in many cases. For instance, this dependence presents challenges in coupling STOVs into a single optical fiber for mode-division multiplexing [33,34], in the angular momentum transfer to particles for optical trapping and manipulation [35], and in the OAM-dependent divergence that accompanies the increase in mode index for optical transmission [36].

Consequently, the concept of perfect optical vortices has been proposed to overcome the above limitations, as their beam radius is independent of the topological charge number [37,38]. This feature introduces an unprecedented paradigm for applications in ultra-secure image encryption [39], high-dimensional quantum teleportation [40], trapping particles [41], and information encoding and transmission [36]. Very recently, Ponomarenko *et al.* introduced the concept of perfect space-time vortices by theoretically incorporating a time lens [42]. In this approach, a Bessel-type STOV is transformed into a perfect STOV through a sequence of an ordinary lens and a time lens. However, implementing a time lens directly in the temporal domain is very challenging, as it requires precise quadratic phase modulation on the picosecond timescale [43].

In this work, we report the experimental generation of perfect spatiotemporal optical vortices, where the ring size in both space and time remains independent of the topological charge. We theoretically and experimentally demonstrate that the proposed perfect STOV wavepackets are the spatiotemporal Fourier transforms of polychromatic Bessel-Gaussian beams based on space-time duality. Experimental results are in excellent agreement with theoretical predictions. Furthermore, we show that the spatiotemporal collision of two perfect STOVs produces STOV lattices, where the number of sub-vortices and symbols can be freely controlled.

## 2. Theoretical Analysis and Numerical Simulations

The perfect STOV wavepacket in the spatiotemporal domain (X−T) can be generated by modulating spatial-spectral field in the $k_x - \omega$ plane. Assume the optical field in the $k_x - \omega$ domain is given by $\psi_0(\rho,\varphi)$, where $(\rho,\varphi)$ are the corresponding polar coordinates, $\rho = \sqrt{k_x^2 + \omega^2}$, and $\varphi = tan^{-1}(\omega/k_x)$. After applying a spiral phase of $e^{-il\varphi}$, a 2D Fourier transform yields the field in the X-T domain, which can be expressed as:

$$\psi(r,\theta) = FT\{\psi_0(\rho,\varphi)e^{-il\varphi}\}, \quad (1)$$

where $r = \sqrt{X^2 + T^2}$, $\theta = tan^{-1}(X/T)$ and FT represents the Fourier transform. The spiral phase and its related optical OAM are conserved after a 2D Fourier transform from the spatial frequency–frequency domain ($k_x - \omega$ plane) to the spatial–temporal domain (X−T plane) [21].

In the spatial domain, a spatially perfect vortex beams can be generated through the Fourier transform of spatial Bessel beams[44]. Motivated by this, to generate perfect STOV wavepackets, we introduce a spatial-spectral coupled polychromatic Bessel-Gaussian beam on the $k_x - \omega$ plane, which is given by: [42]

$$\psi(\rho,\varphi) = \exp\left(-\frac{\rho^2}{w_0^2}\right)J_l(a\rho)e^{-il\varphi}, \quad (2)$$

where $w_0$ is the size of the Gaussian envelope at the beam waist, $J_l$ is the first kind Bessel function with an order of *l*. *a* is the normalization factor that scales the dimensions of the Bessel-Gaussian beam. $e^{-il\varphi}$ is the spiral phase in the spatial-spectral plane. In the experiment, the Fourier spectral field in Eq. (2) can be generated through a recently developed spatiotemporal holographic shaping method that can generate the complex amplitude field distribution with the amplitude distribution being $\exp\left(-\frac{\rho^2}{w_0^2}\right)J_l(a\rho)$ and the phase distribution being $e^{-il\varphi}$. The detail of the holographic shaping process can be found in the Supplementary Section 2. The resulting spatiotemporal field on the X−T plane can be then calculated by a 2D Fourier transform[30,44]:

$$\begin{aligned}\psi(r,\theta) &= FT[\psi(\rho,\varphi)] \\ &= \frac{1}{2\pi}\int_0^{2\pi}\int_0^\infty \exp\left(-\frac{\rho^2}{w_0^2}\right)J_l(a\rho)e^{-il\varphi}e^{i\rho r\cdot\cos(\varphi-\theta)}\rho d\rho d\varphi \\ &\approx i^{l-1}\frac{w_0}{w}\exp\left(-\frac{(r-r_0)^2}{w^2}\right)e^{-il\theta}.\end{aligned} \quad (3)$$

The above equation represents the complex amplitude of perfect STOV with topological charge of $l$ and ring radius $r_0 = af/k_0$. The Gaussian beam waist at the focus is $w = 2f/k_0 w_0$, where $f$ is the focal length of a Fourier lens, $k_0 = 2\pi/\lambda$, $\lambda$ is the wavelength. It can be seen that the ring size of a perfect STOV is independent of topological charge and can be adjusted by changing $a$ and $f$.

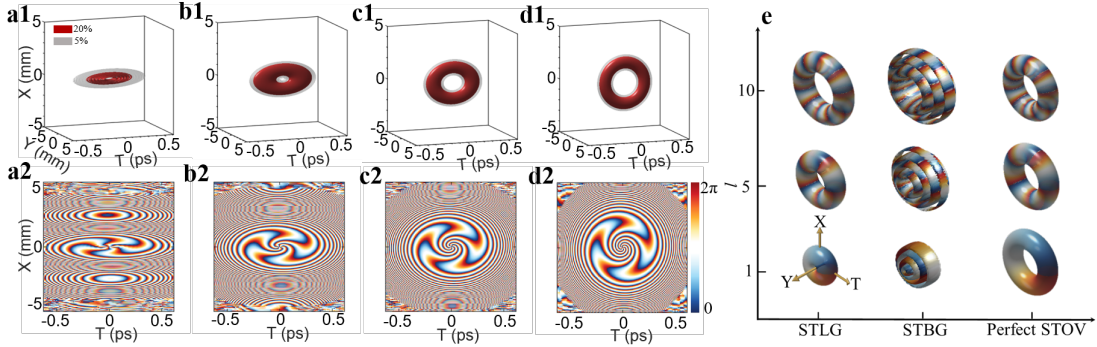

**Fig. 1 Perfect spatiotemporal optical vortices (STOV) wavepackets. a-d** 3D iso-intensity profiles of a perfect STOV of $l = +5$ and corresponding sliced phase patterns (at y = 0) at different locations L = 75, 150, 225, and 300 mm after the spatiotemporal pulse shaper. **e** Comparison of spatiotemporal Laguerre-Gaussian wavepackets (STLG), spatiotemporal Bessel-Gaussian wavepackets (STBG) and perfect STOV with topological charge $l$ of = +1, +5, +10.

Equations (1) to (3) suggest that perfect STOV can be synthesized via spatiotemporal Fourier transform by employing a spatiotemporally coupled polychromatic Bessel-Gaussian seed beam. Figures 1a-d present the numerical simulation results for the generation of perfect STOV wavepackets at different free-space propagation distances. The simulation uses a spatial-spectral Bessel-Gaussian beam with a topological charge of $l = +5$, a spatial width of 0.7 mm, a central wavelength of 1.03 μm, and a spectral width of 10 THz. A perfect STOV wavepacket (Fig.1d) is faithfully generated at a distance L = 300 mm following a 4f pulse shaper. On the other hand, in a medium with anomalous dispersion, the perfect STOV exists only within a finite propagation distance, beyond which it degrades into a Bessel-Gaussian form. (see Supplementary Section 1).

Previous research has demonstrated the generation of spatiotemporal Laguerre-Gaussian and spatiotemporal Bessel-Gaussian wavepackets, where the spatial and temporal diameters are dependent on the topological charge [29,30,32,45]. As shown in Fig. 1e, numerical simulations reveal that the spatial and temporal widths of spatiotemporal Laguerre-Gaussian and spatiotemporal Bessel-Gaussian wavepackets increase with increasing topological charge. In contrast, the spatial and temporal diameters of perfect STOV wavepackets remain constant, regardless of the topological charge. This is one of the main features and advantages of perfect STOVs.

## 3. Experimental Setup

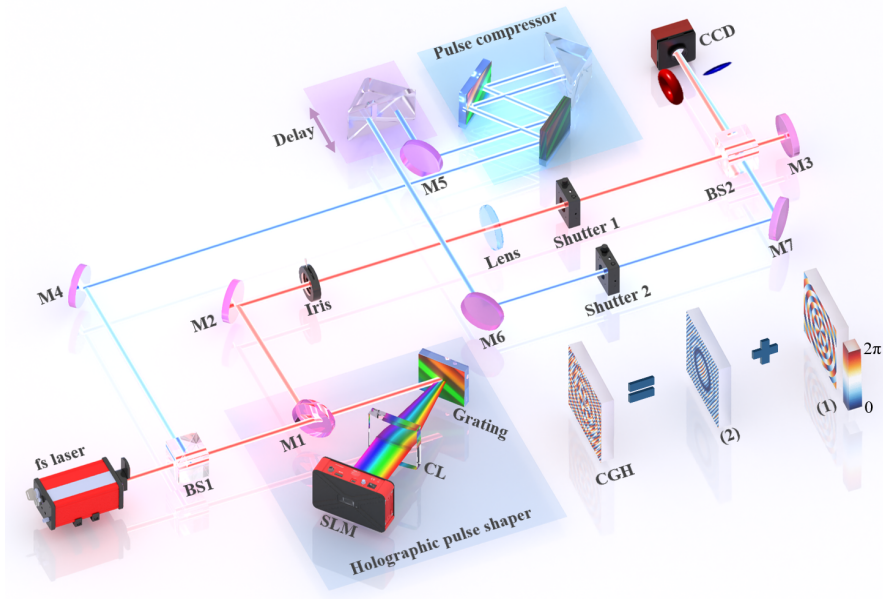

**Fig 2 Experimental setup for synthesizing and characterizing perfect STOV wavepackets.** The setup includes three sections: a holographic pulse shaper, a time delay line system for fully measuring the 3D profile of the generated perfect STOV wavepacket, and a pulse compressor system. The computer-generated hologram (CGH) embedded on the SLM comprises three parts of phase: (1) the phase distribution of a spatial-spectral Bessel-Gaussian mode; (2) a phase-only diffraction grating for controlling the spatial-spectral amplitude modulation. M, mirror; BS, beam splitter.

Figure 2 exhibits the experimental setup for perfect STOV generation and characterization. The mode-locked input laser, with a spectral bandwidth of approximately 20 nm centered at 1030 nm, is split into a probe pulse and an object pulse. The probe pulse passes through a pulse compressor consisting of a pair of parallel gratings and a right-angle prism and is then de-chirped to a Fourier-transform-limited pulse. The object pulse is directed into a folded 2D ultrafast pulse shaper that consists of a reflective grating, a cylindrical lens, and a phase-only reflective SLM (Holoeye GAEA-2, 3840×2160 pixels with a pitch of 3.74μm). The programmable SLM is positioned in the spatial-spectral plane and imprints an intricate phase pattern, as shown in the bottom right corner. This pattern encodes the complex amplitude of the polychrome Bessel-Gaussian beam described in Eq. (2) (The details of the complex amplitude encoding are provided in the Supplementary Section 2). The modulated beam is reflected and recombined by the grating to generate the perfect STOV wavepackets on a specific plane, where a CCD camera is positioned. The spatiotemporal wavepacket is synthesized in the far field via a time-delayed spatial propagation after the grating. To characterize the wavepacket, the probe and generated wavepacket are recombined at the CCD camera with a small tilt angle. By scanning the time delay between the pulses, the information of the object wavepacket is encoded into the fringe pattern, allowing the retrieval of its amplitude and phase profiles.

## 4. Experimental Results and Discussions

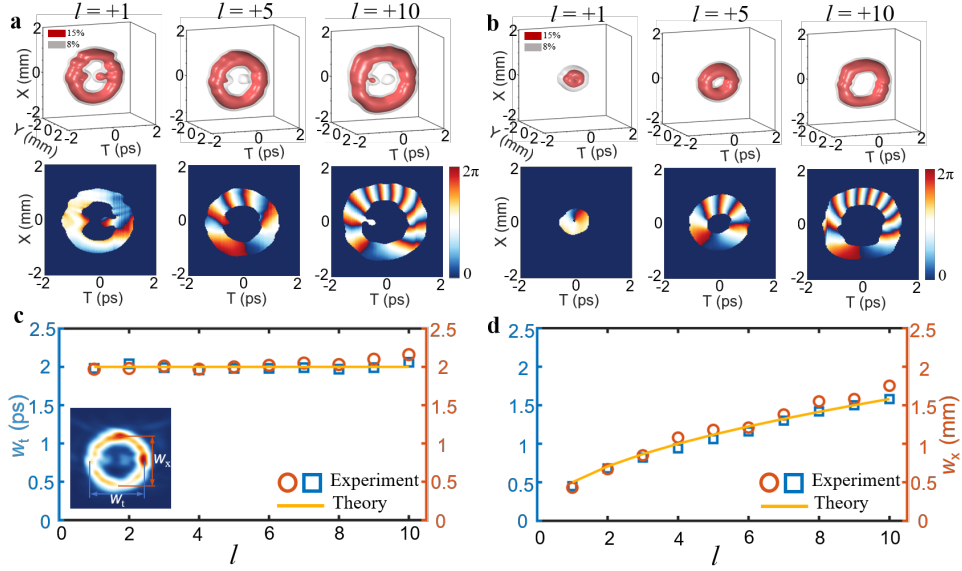

**Fig. 3 Theoretical and experimental results for the generated perfect STOV wavepacket and its comparison with the spatiotemporal Laguerre-Gaussian wavepacket. a b** Reconstructed intensity iso-surface and sliced phases (y = 0 plane) of generated perfect STOVs and spatiotemporal Laguerre-Gaussian wavepackets with topological charge $l$ = +1, +5, +10, respectively. **c d** Dependence of spatial (temporal) diameter on the topological charge of the generated perfect STOVs and spatiotemporal Laguerre-Gaussian wavepackets.

Figure 3a presents the intensity iso-surface and sliced phase patterns of experimentally generated perfect STOV wavepackets with typical topological charges of $l$ = +1, +5, and +10, respectively. The size of the perfect STOV remains constant in spacetime coordinates. To quantitatively assess the properties of perfect STOVs, Fig. 3c plots the variation of spatial ($w_x$, circle marker) and temporal ($w_t$, square marker) diameters as a function of the topological charge. The experimental data (circles and squares) aligns well with the theoretical predictions (solid curves), confirming that the spatial and temporal diameters of perfect STOVs are indeed independent of the topological charge. This experimental verification solidifies the characteristics of perfect STOV wavepackets (see Supplementary Section 3).

To evaluate the performance of perfect STOVs, we conducted an experimental comparison with single-ring spatiotemporal Laguerre-Gaussian [32] of $p$ = 0. Using identical beam parameters and topological charge numbers ($l$ = +1, +5, +10), we generated spatiotemporal Laguerre-Gaussian wavepackets with $p$ = 0. The spatial and temporal diameters of the spatiotemporal Laguerre-Gaussian wavepackets increase significantly with larger topological charge, as shown in Fig. 3b. As seen in Fig. 3d, there is excellent agreement between the experimental data (represented by circles and squares) and the theoretical results (solid curves). These results demonstrate that both the spatial and temporal diameters increase with increasing topological charges.

The properties of the perfect STOV wavepackets are further verified by comparing those with spatiotemporal Laguerre-Gaussian wavepackets. In addition to their unique characteristics of constant sizes, perfect STOVs could offer additional degrees of freedom in controlling their spatial and temporal dimensions. By adjusting the parameter $a$ in Eq. (2), we can freely regulate both the spatial and temporal diameters of perfect STOV wavepackets. To illustrate the advantages of "perfect" in space-time, we examined two cases, following the method outlined in Ref. [45], where two perfect STOVs with applied linear phase coefficients $k_t = -1.5$ ps and $k_t = +1.5$ ps move in opposite directions along the time axis.

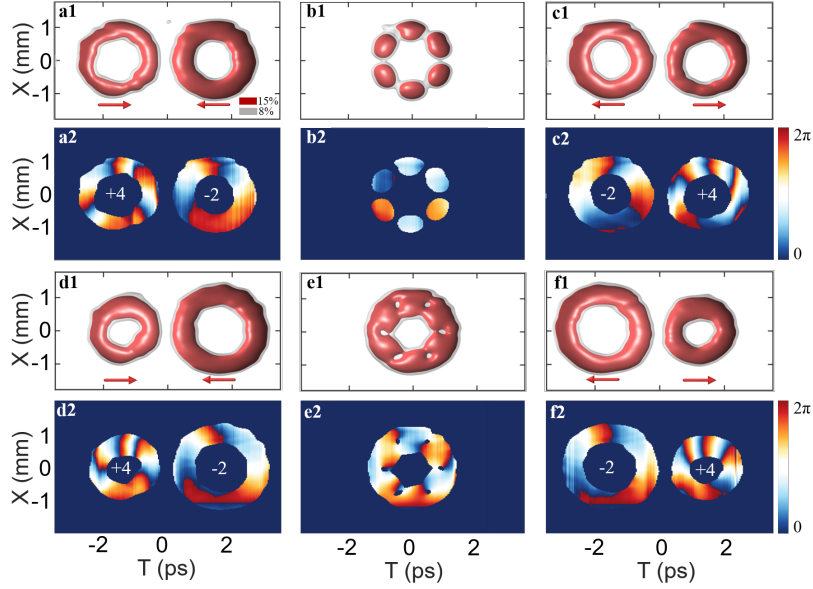

**Fig. 4 Spatiotemporal collision of perfect STOVs with time-varying transverse OAM. a1 – c1** Measured iso-intensity profile of the collision of two perfect STOVs with the same spatial and temporal diameters and different topological charges of $l_1 = +4$, $l_2 = -2$. **a2 – c2** Retrieved phase distribution in the meridional plane corresponding to a1 – c1. **d1 – f1** Measured iso-intensity profile of the collision of two perfect STOVs with different spatial and temporal diameters and different topological charges of $l_3 = +4$, $l_4 = -2$. **d2 – f2** Retrieved phase distribution in the meridional plane corresponding to d1 – f1. Linear phases along the spectral direction with opposite signs are applied on the left/right side of the input light field to advance/delay input wavepackets in the corresponding time domain. The phase is expressed as $\varphi(\omega) = k_t(\omega - \omega_0)$. From **a** to **c** (**d** to **f**), the linear phase coefficient $k_t$ changes from −1.5 ps to 1.5 ps. The arrow direction indicates the direction in which the perfect STOV is shifted in the time domain.

In the first case, as shown in Fig. 4a1 and a2, the initial two perfect STOV wavepackets separated in time are generated first by applying linear phases with opposite slopes to perfect STOV wavepackets with the same spatial and temporal diameters ($a_1 = a_2 = 5$mm$^{-1}$) and $l_1 = +4$ and $l_2 = -2$. Then, we accelerate and delay the two wavepackets to 0 ps respectively in time, so that the two wavepackets collide in the space-time domain, forming a petal-like interference pattern, as shown in Fig.4b1, and the number of petals is $N = |l_2 - l_1|$. In the second case, as shown in Fig. 4d and f, two perfect STOV wavepackets of different spatial and temporal diameters ($a_3 = 4$mm$^{-1}$ and $a_4 = 6$mm$^{-1}$) and $l_3 = +4$ and $l_4 = -2$. When these two perfect STOV wavepackets collide completely (Fig.4 e1), it is clear that there are some dark dots at the contact point of the two perfect STOV wavepackets. As shown in Fig.4e2, the phase pattern of the superposition of the two perfect STOV wavepackets shows that spatiotemporal optical vortex with $l = -1$ exists at each dark dot. The number of spatiotemporal phase singularity is $N = |l_4 - l_3|$. Furthermore, their signs are easily determined by the sign of $N = l_4 - l_3$. Therefore, the collision of these two perfect STOV wavepackets with different spatial and temporal diameters leads to the generation of an STOV lattice with controllable numbers and signs of vortices. Therefore, the perfect property and radial degrees of freedom of perfect STOV wavepackets can be used to flexibly generate more rich and complex spatiotemporal structure light fields.

## 5. Conclusions

In conclusion, perfect spatiotemporal optical vortex with spatiotemporal sizes independent of the topological charges is generated experimentally by the spatiotemporal Fourier transform of the polychrome Bessel-Gaussian beam. The properties of the perfect STOVs are verified through comparison with spatiotemporal Laguerre-Gaussian wavepackets. Furthermore, we show the spatiotemporal collision of two perfect STOV wavepackets by introducing a linear phase in the time dimension, where the spatiotemporal collision of two perfect STOVs with different diameters leads to the phenomenon of spatiotemporal vortex reconnection, resulting in STOV lattice consisting of spatiotemporal sub-vortices with controllable number and sign. These perfect spatiotemporal optical vortices may be used to facilitate optical communications, particle trapping, and tweezing among other potential applications.

**Funding.** We acknowledge financial support from National Natural Science Foundation of China (NSFC) [Grant Nos. 12434012 (Q.Z.) and 12474336 (Q.C.)], the Shanghai Science and Technology Committee [Grant Nos. 24JD1402600 (Q.Z.) and 24QA2705800 (Q.C.)], National Research Foundation of Korea (NRF) funded by the Korea government (MSIT) [Grant No. 2022R1A2C1091890], and Global - Learning & Academic research institution for Master's·PhD students, and Postdocs (LAMP) Program of the National Research Foundation of Korea(NRF) grant funded by the Ministry of Education [No. RS-2023-00301938]. Q.Z. also acknowledges support by the Key Project of Westlake Institute for Optoelectronics [Grant No. 2023GD007].
**Disclosures.** The authors declare no conflicts of interest.
**Data availability.** Data underlying the results presented in this paper are not publicly available at this time but may be obtained from the authors upon reasonable request.
**Supplemental document.** Please see **Supplement 1** for supporting content.